\documentclass{llncs}

\usepackage{a4wide}

\usepackage[french]{babel}
\usepackage[utf8]{inputenc}

\usepackage{lmodern}
\usepackage[T1]{fontenc}
\usepackage{xspace}
\usepackage{microtype}

\begin{document}

\title{Faire levier sur les architectures logicielles pour guider et
  vérifier le développement d'applications SCC\thanks{Ces deux pages
    résument l'article \emph{``Leveraging Software Architectures to Guide
    and Verify the Development of Sense/Compute/Control
    Applications''} publié à ICSE ’11}}
\titlerunning{Faire levier sur les architectures logicielles}

\author{Damien Cassou\inst{1} \and Emilie Balland\inst{2} \and Charles
  Consel\inst{2} \and Julia Lawall\inst{3}}
\authorrunning{Damien Cassou et al.}
\tocauthor{Damien Cassou, Emilie Balland, Charles Consel, and Julia
  Lawall}
\institute{%
Software Architecture Group, Hasso-Plattner-Institut, Potsdam, Germany,\\%
\and%
INRIA/University of Bordeaux, France,%
\and%
DIKU, University of Copenhagen, Denmark%
 }

\maketitle              % typeset the title of the contribution

\begin{abstract}
  Une architecture logicielle décrit la structure d'un système
  informatique en spécifiant ses composants et leurs interactions.
  Projeter une architecture logicielle sur une implémentation est une
  tâche reconnue difficile. Un élément crucial de cette projection est
  la description architecturale des interactions entre les composants.
  La caractérisation de ces interactions peut être plutôt abstraite ou
  très concrète, fournissant plus ou moins de support de programmation
  et de possibilités de vérifications statiques.

  Nous explorons un point dans l'espace de conception entre les
  spécifications abstraites et concrètes des interactions de
  composants. Nous introduisons la notion de \emph{contrat
    d'interactions} qui exprime les interactions autorisées. Cette
  déclaration architecturale permet la génération de support de
  programmation qui assure la conformité entre l'architecture et
l'implémentation, et favorise diverses vérifications. Nous instancions
  notre approche sur un langage de description d'architectures pour
  les applications \emph{Sense/Compute/Control} et décrivons les
  stratégies de compilation et de vérification associées.
\end{abstract}

\section{Introduction}

Une application \emph{Sense/Compute/Control} (SCC) est une application
qui interagit avec un environnement extérieur. Ces applications se
retrouvent dans des domaines comme la domotique, la robotique et
l'informatique autonome. Développer une application SCC est complexe
car l'implémentation doit prendre en compte l'interaction avec
l'environnement. De plus, la correction est essentielle puisque un
changement dans l'environnement peut avoir des conséquences
irréversibles.

Une application SCC peut être définie suivant un style architectural
comprenant quatre types d'éléments organisés en couches : (1) les
\emph{sources}, en bas, obtiennent les informations de l'environnement
; (2) les \emph{opérateurs de contexte} traitent ces informations ;
(3) les \emph{opérateurs de contrôle} utilisent ces informations
raffinées pour contrôler (4) les \emph{actions}, en haut, qui
impactent finalement l'environnement. Projeter une architecture
logicielle ayant un tel niveau d'abstraction vers une implémentation
et maintenir cette projection sont des tâches reconnues difficiles.

Dans cet exposé nous proposons une approche pour lier architecture et
implémentation qui vise les applications SCC. Cette approche introduit
la notion de \emph{contrat d'interactions} permettant à un architecte
de déclarer quelles sont les interactions qu'un élément de
l'architecture a le droit de réaliser (Section~\ref{sec:contrats}).
Cette notion de contrat d'interactions est dédiée au style
architectural SCC dans le sens où un contrat d'interactions ne peut,
syntaxiquement, décrire que les interactions autorisées par le style.
Les contrats d'interactions sont utilisés pour générer un support de
programmation qui va guider le travail d'implémentation par les
développeurs tout en maintenant la conformité avec l'architecture
(Section~\ref{sec:implementation}). L'architecte peut aussi utiliser
les contraintes exprimées par les contrats d'interactions pour
vérifier un ensemble de propriétés allant au delà de la conformité
(Section~\ref{sec:analyses}).

\section{Contrats d'interactions}
\label{sec:contrats}

Le but d'un contrat d'interactions est de décrire les interactions
autorisées d'un opérateur au sein d'une application SCC. Ce contrat
d'interactions est un triplet constitué des informations suivantes :
\emph{la condition d'activation} permet d'indiquer quelles sont les
interactions capables d'activer l'opérateur ; \emph{les données
  requises} permettent d'indiquer les interactions supplémentaires
autorisées pour chaque condition d'activation ; \emph{les actions à
  entreprendre} permettent d'indiquer la réponse appropriée à chaque
activation (émission d'une information pour un opérateur de contexte
ou commande d'une action pour un opérateur de contrôle). En résumé,
les contrats d'interactions guident le travail de l'architecte en
lui proposant un cadre de spécification dédié au style SCC.

\section{Support de programmation}
\label{sec:implementation}

Nous avons intégré les contrats d'interactions dans DiaSpec, un
langage de description d'architectures dédié aux applications SCC. À
partir d'une architecture en DiaSpec, un générateur de code produit un
\emph{framework} de programmation Java dédié.
Ce \emph{framework} de programmation généré contient une \emph{classe
  abstraite} pour chaque élément de l'architecture. Cette classe
abstraite générée contient des méthodes pour faciliter
l'implémentation des éléments ainsi que des déclarations de méthodes
abstraites permettant d'implémenter la logique applicative.
Implémenter un élément DiaSpec nécessite donc de créer une sous-classe
de la classe abstraite générée correspondante. En conséquence, dans
cette approche, un architecte peut changer l'architecture et générer
un nouveau \emph{framework} de programmation sans écraser le code des
développeurs. Les changements dans l'architecture qui ont un effet sur
le code déjà implémenté sont révélés par le compilateur Java assurant
par là la conformité de l'implémentation avec l'architecture.

Chaque contrat d'interactions d'un opérateur se projette vers la
déclaration d'une méthode abstraite dans la classe abstraite générée
correspondante à l'opérateur. En particulier, (1) la condition
d'activation influence le nom de la méthode abstraite ainsi que son
premier paramètre ; (2) les données requises se projettent vers
autant de paramètres représentant des fonctions permettant d'exécuter
l'interaction supplémentaire ; (3) les actions à entreprendre se
projettent vers un ou plusieurs paramètres supplémentaires ainsi que
vers le type de retour de la méthode.

Le \emph{framework} de programmation est généré de façon à guider les
développeurs dans l'implémentation de l'application ainsi qu'à les
limiter à ce que l'architecture autorise. En particulier, un contrat
d'interactions se projette vers une méthode abstraite qui fournit en
paramètre tout ce qui est nécessaire à l'implémentation de la logique
applicative de l'opérateur. De plus, la déclaration de cette méthode
impose au développeur de respecter les contraintes de l'architecture,
assurant ainsi la conformité entre l'architecture et l'implémentation.

\section{Support de vérification}
\label{sec:analyses}

Les contrats d'interactions rendent explicites des informations sur le
flot de données et permettent des vérifications statiques. Par
exemple, avec les contrats d'interactions, il est possible de savoir
au moment de la conception tous les opérateurs qui seront
éventuellement activés par la publication d'une information par une
source. De plus, notre stratégie de génération assure que ces
propriétés seront préservées au niveau de l'implémentation.

Les contrats d'interactions permettent aussi de vérifier des
\emph{invariants d'interactions} qui sont des propriétés vérifiées à
tout moment de l'exécution. Nous caractérisons la progression d'une
application SCC par son flot de données et utilisons la logique
temporelle linéaire (LTL) pour définir ces invariants. Pour vérifier
ces invariants, une architecture DiaSpec est automatiquement traduite
en un modèle pour le \emph{model checker} SPIN. Si un invariant n'est
pas satisfait, SPIN donne un contre exemple sous la forme d'une trace
d'exécution qui guide l'architecte dans la correction de son
architecture. Cet exemple montre que les contrats d'interactions
rendent explicites les concepts clés du style architectural SCC et
donc facilitent les analyses sur le flot de données.

\section{Conclusion}

Nous avons introduit la notion de contrat d'interactions exprimant les
interactions autorisées au sein d'une architecture SCC. Les contraintes
exprimées par ces contrats d'interactions permettent des vérifications et
guident l'implémentation de l'architecture, tout en assurant la conformité.

\end{document}